# A thought experiment on quantum mechanics and distributed failure detection

*M. C. Little*

One of the biggest problems in current distributed systems is that presented by one machine attempting to determine the liveness of another in a timely manner. Unfortunately, the symptoms exhibited by a failed machine can also be the result of other causes, e.g., an overloaded machine or network which drops messages, making it impossible to detect a machine failure with certainty until that machine recovers. This is a well understood problem, and one which has led to a large body of research into *failure suspectors*: since it is not possible to *detect* a failure, the best one can do is *suspect* a failure and program accordingly. However, one machine's suspicions may not be the same as another's; therefore, these algorithms spend a considerable effort in ensuring a consistent view among all available machines of who is suspected of being failed. This paper describes a thought experiment on how quantum mechanics may be used to provide a failure detector that is guaranteed to give both accurate and instantaneous information about the liveness of machines, no matter the distances involved.

## 1    Problem statement

With the advent of the Web, distributed computing has become a pervasive technology. Individual computers frequently communicate with each other in order to perform work, without knowing where they are physically located. Typically the only adverse effect that distribution has is on the latency with which responses occur. However, the apparent ease with which these computations take place belies the fact that distribution poses problems not typically seen in centralised systems: failure of machines can occur independently, possibly leading to inconsistencies within applications or the inability to perform work.

Therefore, fault-tolerant techniques attempt to mask component (machine) failures from applications, and typically rely upon the fact that such failures can be detected [1][2]. An *ideal failure detector* is one which can allow for the unambiguous determination of the liveliness of an *entity*, (where an entity may be a process, machine etc.,), within a distributed system. However, as shown in [3][4], guaranteed detection of failures in a finite period of time is not possible because it is not possible to differentiate between a failed system and one which is simply slow in responding.

Current failure detectors use timeout values to determine the liveness of entities: for example, if a machine does not respond to an "are-you-alive?" message within a specified time period, it is *assumed* to have failed. If the values assigned to such timeouts are wrong (e.g., because of network congestion), incorrect failures may be assumed, potentially leading to inconsistencies when some machines "detect" the failure of another machine while others do not [2]. Therefore, such timeouts are typically assigned given what can be assumed to be the worst case scenario within the distributed environment in which they are to be used, e.g., worst case network congestion and machine load. However, distributed systems and applications rarely perform exactly as expected from one execution to another. Therefore, fluctuations from the worst case assumptions are possible, and there is always a finite probability of making an incorrect failure detection decision.

Given that guaranteed failure detection based upon timeouts is not possible, there has been much work on *failure suspectors* [3]: a failure suspector works by realising that although guaranteed failure detection is impossible, enforcing a decision that a given entity may have failed on to other, active entities is possible. Therefore, if one entity assumes another has failed, a protocol is executed between the remaining entities to either agree that it will be assumed to have failed (in which case it is *excluded* from the system and no further work by it will be accepted) or that it has not failed: the fact that one entity thinks it has failed does not mean that all entities will reach the same decision. If the entity has not failed and is excluded then it must eventually execute another protocol to be recognised as being alive.

The advantage of the failure suspector is that all correctly functioning entities within the distributed environment will agree upon the liveness of another suspected failed entity. The disadvantage is that such failure suspection protocols are heavy-weight, typically requiring several rounds of agreement. In addition, since suspected failure is still based upon timeout values, it is possible for non-failed entities to be excluded, thus reducing (possibly critical) resource utilisation and availability.

Some applications can tolerate the fact that failure detection mechanisms may occasionally return an incorrect answer. However, for other applications the incorrect determination of the liveliness of an entity may lead to problems such as data corruption [2], or in the case of mission critical applications (e.g., aircraft control systems or nuclear reactor monitoring) could result in loss of life.

In this paper we shall describe a thought experiment on how it may be possible to use quantum mechanics to enable the design of an ideal failure detector: such a failure detector will impose negligible overheads on applications, is guaranteed to never give an incorrect answer, and allows instantaneous detection of component failures *no matter how distributed the system is*. Unfortunately, at present we cannot describe an implementation of the failure detector: the necessary technology does not yet exist to the best of our knowledge. Therefore, all we can present is an indication of how things *might* eventually work.

## 2   Principles of quantum mechanics

In order to describe the idea behind our failure detector, it is first necessary to understand some basic principles of quantum mechanics. Prior to the 20th Century it was believed that electrons behaved like particles. However, in the early 20th Century it was discovered that electrons, light and sub-microscopic particles[1] possess properties of both waves and particles (*wave-particle duality*) [5]. The classic experiment to illustrate this involves an electron gun positioned behind a grating which has two slits in it, as shown in Figure 1. All of the electrons emitted from the gun have (nearly) the same energy. Behind the grating is a thin plate which serves as a stop and on this is a moveable electron detector which counts the average number of electrons arriving at a specific point: essentially the probability that an electron is incident at a particular point.

---

[1] In fact de Broglie showed that all particles possess both wave and particle duality.

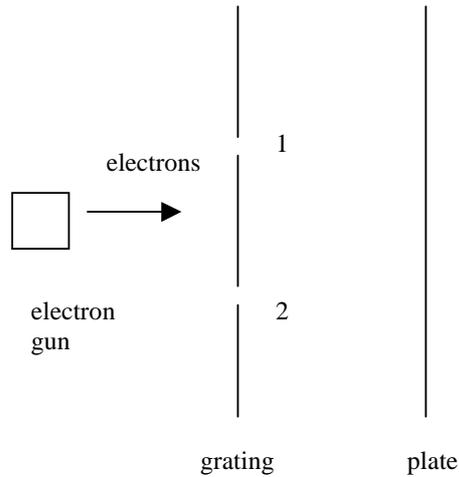

**Figure 1: Electron experiment.**

If electrons were particles as originally assumed, it would also be reasonable to assume that a particle which is incident on the grating will either pass through slit 1 or slit 2 (like shooting a shotgun at the holes, the individual pellets will go through one slit or another). In which case the probability of an electron being incident at a specific point on the plate should be the sum of the probability of the electron going through slit 1 *and* the probability of it going through slit 2. Since a particle cannot pass through both slits, the expected pattern would be as shown in Figure 2.

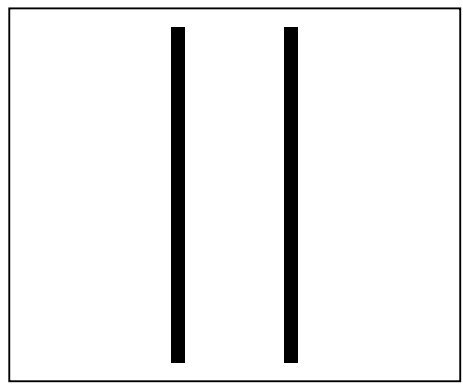

**Figure 2: Particle diffraction pattern.**

However, the actual pattern measured at the plate is shown below in Figure 3:

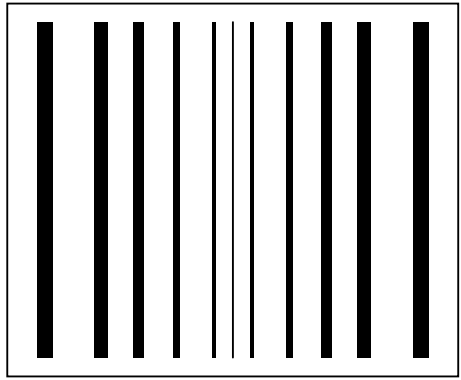

**Figure 3: Electron diffraction pattern.**

The alternating pattern of dark and light bands is similar to that which occurs when water waves are incident on a grating: a water wave incident on the grating produces two sources of waves, one for each slit; these waves then interfere constructively (the dark bands) and destructively (the light bands). The probability distribution measured at the plate therefore implies that the electrons are undergoing interference. The electrons arrive at the plate in discrete "lumps" (packets or quanta), but the probability of arrival is distributed like the intensity of a wave.

It would be natural to try to determine the exact path an electron travels by placing a strong light source between the grating and the plate: electric charges scatter light, so when an electron passes through, however it does so, it will scatter some light to a detector so we can "see" the path it takes. However, when the light is so positioned, the detector measures a different probability distribution at the plate than previously detected: it now detects the pattern shown in Figure 2! The distribution is now the sum of the probability of the electron passing through slit one and the probability of it passing through slit two, i.e., the electrons now behave as particles and not waves. Interestingly, if the light is turned out, the original probability distribution returns.

Therefore, we must conclude that looking at the electrons changes their behaviour! This is because the photons of light are affecting the electrons [5]. In fact it is impossible to arrange an experiment so that we can tell which hole an electron goes through without affecting the probability pattern at the plate. This forms the basis of Heisenberg's Uncertainty Principle: "the more precisely the position is determined, the less precisely the momentum is known in this instance, and vice versa" [6].

Another interesting feature of the quantum properties of electrons can be illustrated by restoring the experiment to its original form (i.e., removing the light source) and slowly reducing the energy at the electron gun to control the number of electrons it emits. The wave-like probability distribution remains, as expected. However, when we eventually control the gun to emit a single electron, the probability distribution remains the same, i.e., it would appear as though the single electron is interfering with itself! What this and other experiments imply, is that if a quantum "particle" is free to travel through multiple routes it actually travels through all possible paths! Only if we look at it do we force it to go through a specific route: each path travelled will have a specific probability of being the one chosen. The particle is said to be in a state of *superposition* of all possible states [7][8].

This property of quantum mechanics, known as *quantum entanglement*, enables two "particles", described by a single wave function, to be arranged such that an experiment performed on one can give information about the other, instantaneously, no matter how far apart the two particles are [9][10]. The rest of this paper builds upon this principle to construct a precise failure detector.

If the collapse of the wave functions describing one of the entangled "particles" occurs this causes the collapse of the entangled wave function as a while; this happens immediately, no matter how far apart the two wave functions may be. Einstein was particularly troubled by this property of quantum mechanics since it breaks the special theory of relativity: information about the specific location of the electron is communicated faster than the speed of light between the two probable locations [5][9].

## 3    The quantum failure detector

Given the brief description of some of the principles of quantum mechanics, and quantum entanglement in particular, this section shall describe our thought-experiment failure detector and how it might be used to guarantee detection of machine failures. To simplify the discussion, we shall concentrate on a simple distributed system consisting of only two machines; however, the principles to be described should be extendable to any number of machines separated by arbitrary distances.

In the Quantum Failure Detector, we use a quantum particle (e.g., an electron) and pass it through a grating as illustrated in Figure 1, such that the particle's wave function is described by two components: the entangled "particles". As show in Figure 4, for simplicity we assume that there is a single device which is responsible for creating this entangled wave function and passing it to the nodes in the distributed system. In reality each machine may possess its own such device and be responsible for generating and diffusing these particles to other machines in the distributed system when it becomes active.

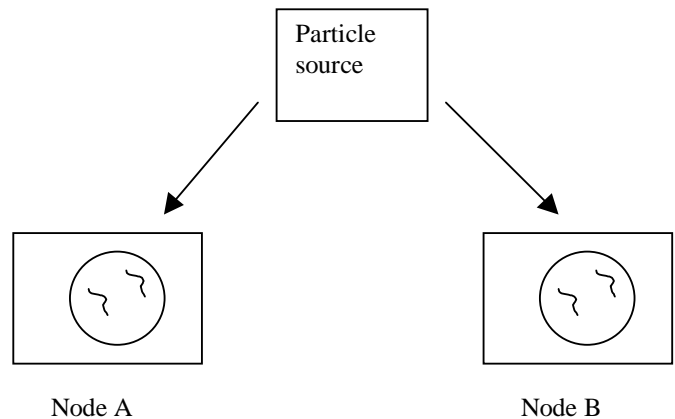

**Figure 4: Quantum entangled particle creator and distribution.**

Each node in the system is thus given a component of the entangled wave function, which it must maintain in that quantum state, i.e., the particles must remain in their superposed state. How this occurs is beyond the scope of this paper and may depend upon the type of particle being used, e.g., an *electron trap*.

Each node therefore maintains 2 different wave functions, one for itself, and one for the other node in the system, both components of their respective entangles wave functions. For the failure detector to work, as long as a machine remains alive it must maintain both of these wave functions in their *superposed* states, i.e., it must do nothing to cause the respective wave function to collapse. If a node fails, the Quantum Failure Detector requires that the failing node perform an experiment on its own copy of the entangled particle in order to cause it to collapse from its superposed state, i.e., force a specific state to be associated with the particle – essentially determine which slit of the original grating the particle passed through.

When this happens, the corresponding entangled "particle" held by the other machine will also instantaneously collapse onto a different specific state. Since in the macroscopic world a particle can only be in one place or another, either the particle exists at the failed machine or it exists at the other machine. This state can be measured in a number of ways; for example, as illustrated in Figure 5 we assume that the "particle" wave at node A vanishes causing its counterpart at node B to produce the particle pattern we saw in Figure 2 (prior to this it would produce the wave pattern illustrated by Figure 3). Obviously the act of "examining" the particle must not cause a superposed particle to collapse onto a specific state, otherwise false failure detection will occur.

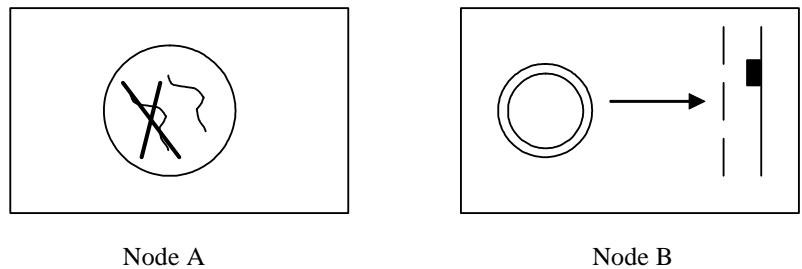

Node A                                          Node B

**Figure 5: Quantum failure detector.**

As soon as the particle no longer exists within a superposed state, the remaining machine knows that the machine the particle represents has failed, since this is the only event that can cause the superposed probability function to collapse. Importantly, this knowledge is guaranteed to be correct, and will be seen instantaneously by the surviving machine no matter how far apart they may be separated.

For example, if A needs to determine the current liveness status of node B, it need only examine its local node B particle: no direct communication between machines is required as this will implicitly occur within the quantum world faster than we can do in the macroscopic world. As long as the particle remains in a superposed state (e.g., it still causes a wave-like interference distribution when incident on a diffraction grating), then node A can know with certainty that node B is still alive.

## 3.1 Extension

For simplicity the previous discussion was limited to a two node distributed system. If there were $n$ machines in the distributed system, then each machine would need to exchange entangled wave function components with every other machine in the system; this could be accomplished by having n slits in the grating at the particle source in Figure 5. Therefore, each machine would have $n$ entangled particles, 1 for itself, and $n$-1 for the other machines. Importantly, the quantum entanglement effect will allow all available machines to observe the failure of another machine at the same time, i.e., there can be no ambiguity between different machines in the environment.

## 4  Conclusions

We have described the problem encountered in classic distributed computing systems of one machine attempting to accurately determine the liveness of another machine in a timely manner. Such a determination is impossible to make without waiting for the failed machine to eventually recover, since an overloaded machine or network can make a machine appear to be failed. Hence failure suspector protocols have been developed to ensure that one machines suspicions about the availability of another can be forced onto other available machines. However, these protocols are heavyweight and can result in the exclusion of available machines.

In this paper we have given a thought experiment which attempts to show how quantum entanglement can be used to pass information about machine failures instantaneously between machines, regardless of the distances that separate them. This information flow is also accurate. We have outlined how a possible Quantum Failure Detector may use this effect to provide an ideal failure detector that can give accurate and instantaneous information about a machine's liveness.

## 5  Acknowledgements

I would like to thank Doctor Alan Dickinson from the Department of Physics at Newcastle University for commenting on earlier drafts of this paper, and clarifying some of the issues involved. I would also like to thank Professor Santosh Shrivastava for the many interesting and stimulating discussions about failure detection we have had over the years.